\documentstyle[12pt]{article}

         \def\I{{\rm I}}
         \def\d{{\rm d}}
         \def\ds{\displaystyle}
         \def\s{\sigma}
         \def\l{\lambda}
         \def\la{\label}
         \def\r{\ref}
         \def\m{\mu}
         \def\be{\begin{equation}}
         \def\bea{\begin{eqnarray}}
         \def\ee{\end{equation}}
         \def\eea{\end{eqnarray}}
\begin{document}
\begin{titlepage}
\vspace*{20mm}
\begin{center} {\Large \bf A two--parametric family of asymmetric exclusion
                processes\\
\vskip 0.35cm
and its exact solution}\\
\vskip 1cm
\centerline {\bf
M. Alimohammadi$ ^{a}$ \footnote {e-mail:alimohmd@netware2.ipm.ac.ir},
V. Karimipour$ ^{b,c}$ \footnote {e-mail:vahid@netware2.ipm.ac.ir}
M. Khorrami$ ^{c,d}$ \footnote {e-mail:mamwad@netware2.ipm.ac.ir}}

\vskip 1cm
{\it $^a$ Department of Physics, University of Teheran, North Karegar,} \\
{\it Tehran, Iran }\\
{\it $^b$ Department of Physics, Sharif University of Technology, }\\
{\it P.O.Box 11365-9161, Tehran, Iran }\\
{\it $^c$ Institute for Studies in Theoretical Physics and Mathematics,}\\
{\it P.O.Box 19395-5746, Tehran, Iran}\\
{\it $^d$ Institute for Advanced Studies in Basic Sciences,}\\
{\it P.O.Box 159,Gava Zang,Zanjan 45195,Iran.}\\
\end{center}

\vskip 1.5cm
\begin{abstract}
\noindent
A two--parameter family of asymmetric exclusion processes for particles on a
one-dimensional lattice is defined.  The two parameters of the model control
the driving force and an effect which we call pushing, due to the fact that
particles can push each other in this model. We show that this model is
exactly solvable via the coordinate Bethe Ansatz and show that its
{\it N}-particle
$S$-matrix is factorizable. We also study the interplay of the above effects in
determining
various steady state and dynamical characteristics of the system.
\end{abstract}


\end{titlepage}
\newpage
\section{Introduction}

The asymmetric simple exclusion process (ASEP) is the simplest and most studied
model of interacting particle systems in one dimension
(see [1-5] and references therein). This
model can be related via suitable mappings to many different physical
 models ranging from interface
growth [6,7] to problems of traffic flow [8-11]. In this paper,
however we would like to look at it only as a model
for a collection of random walkers interacting with each other by simple
exclusion.
Each particle hops with rate $ R(L)$ to it's right (left) neighboring site if
this site is empty, otherwise it stops. When the hoping rates
$ R $ and $ L $ are equal,
one deals with the symmetric exclusion process. This case, models diffusion
of particles on a one dimensional line in the absence of a driving force.
The other extreme case (i.e. $L$=0), is usually called the totally
asymmetric exclusion process (TASEP).\\
The asymmetric exclusion process has been extensively studied in the past few years as a prototype
of a one dimensional system far from equilibrium for which some exact results
can be obtained. Among the many aspects of this problem which have been studied,
one can mention the mean field solution of the steady state and the
phase structure [12,13], the exact steady state of the open [14], and closed
chain [15],
the effect of impurities [16-18], exact calculation of some dynamical
properties
[19],the effect of different kinds of updatings [20-21], and finally the exact
calculation
of conditional probabilities via the coordinate Bethe Ansatz [22].\\
In ref.[23] we considered the TASEP and added a new element into this
process, namely the possibility that a particle pushes the particles in front
of it with a rate depending on the number of these particles.
Being interested in exactly solvable models, we found that
if a particle can push
a collection of {\it n}-adjacent particles in its immediate neighborhood
with a rate given by
\be\la{1}
r_n={1\over{1+{\l\over\m}+\cdots +\left({\l\over\m}\right)^n}},
\ee
where $ 0\leq \m = 1-\l \leq 1$, then the problem allows an exact solution via
the coordinate Bethe Ansatz.
The fact that $ r_n$ decreases with $ n $  is physically natural,
although its functional
form may seem peculiar. This particular form is
dictated
by our demand of the exact solvability of the model.Varying the
parameter $ \m $ from
0 to 1, one can then smoothly interpolates between the TASEP
and the so called
drop-push model [24].\\
{\bf Remark}: Here, by an exact solution we mean an exact determination of the
time dependence of $n$-particle conditional probabilities on an infinite lattice.
We also show that the steady state of the system on a ring is one in which
all of the configurations have equal weights. The steady state of the system on an
open chain is not known at present. Incidentally, the technique of matrix
product ansatz does not work for this latter problem since the Hamiltonian of
the process is highly nonlocal.\\
In the present work we consider the partially asymmetric case and study
the combined effect of pushing and driving. Again we demand exact solvability
and follow the strategy of our previous work [23]. The basic objects we are
interested in are the probabilities  $P(x_1, x_2,\cdots ,x_N;
t\vert y_1, y_2,\cdots ,y_N) $ for finding at time t, particle
1 at point $ x_1 $, particle 2 at point $ x_2$,$\cdots$, given their initial
positions at points $ y_1, y_2,\cdots $, respectively.
For notational convenience, in the following we suppress the initial
coordinates whenever no confusion can arise.
The physical region
for the coordinates is $ x_1 < x_2 < ... < x_N $.
If one wants to take into account a type of pushing effect, one should write a
large number of equations depending on which and how many of the
particles are adjacent to each other. The number of cases and hence
the number of equations grows rapidly as the number of particles
increases. The basic idea is to write one single equation supplemented by a
particular boundary condition and to see what kind of pushing effect emerges.
The master equation is
\bea\la{2}
{\partial\over \partial t}P(x_1, x_2,\cdots ,x_N;t)&=&R[ P(x_1-1, x_2,
\cdots ,x_N;t) +\cdots + P(x_1, x_2,\cdots ,x_N-1;t)]\cr
&& +L[ P(x_1+1, x_2,\cdots ,x_N;t) +\cdots
+P(x_1, x_2,\cdots ,x_N+1;t)]\cr &&- N P(x_1, x_2,\cdots , x_N;t ).
\eea
In the following we re-scale time so that $ R + L = 1 $.
We assume that this equation holds in the whole physical region. When any of
the coordinates of the left hand side are adjacent (say $ x_{i+1} = x_i + 1 $)
, points on
the boundary of the physical region appear on the right hand side of (2). We
fix the value of these terms by the following boundary condition:
\be\la{3}
P(x,x;t)=\l P(x, x+1;t)+\m P(x-1,x;t),
\ee
where for simplicity we have suppressed all the other coordinates.
This idea is most apparent in sectors of low particle number.
 For example, in the
two particle sector, combination of (\r{2}) and (\r{3}) yields for
adjacent particles
\bea\la{4}
{\partial\over \partial t}P(x, x+1)&=&R[ P(x-1, x+1
) +\m P(x-1, x)- (1+\m )P(x, x+1)]\cr
&& +L[ P(x, x+2)+\l P(x+1, x+2) - ( 1+\l)P(x, x+1)],
\eea
which means that a particle hops freely to the right and left with rates
$ R $ and $ L $  and pushes its neighboring particle to the right and
left with rates
$ R\m $ and $ L\l$, respectively.
We will show that in the general case, equations (\r{2}) and (\r{3}) imply
that the following processes :
\be\la{5}
\underbrace {11\cdots 1}_{n+1}\emptyset \longrightarrow
\emptyset\underbrace {11\cdots 1}_{n+1},
\ee
\be\la{6}
\emptyset\underbrace {11\cdots 1}_{n+1},
\longrightarrow \underbrace {11\cdots 1}_{n+1}\emptyset
\ee
occur respectively  with rates $ r_n $ and $ l_n $ given by
\be\la{7}
r_n=R{1\over{1+{\l\over\m}+\cdots +\left({\l\over\m}\right)^n}},
\ee
and
\be\la{8}
l_n=L{1\over{1+{\m\over\l}+\cdots +\left({\m\over\l}\right)^n}}.
\ee
{\bf Remark:} Interchanging particles and holes, this model can be seen
to be equivalent
to a model in which each particle does not only hops to its immediate
neighboring sites but to any other
vacant site, with rates depending on the hopping distance; i.e., in the
equivalent model the following processes
occur with rates $ r_n $  and $ l_n $, respectively
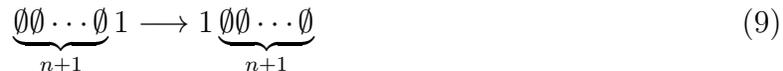
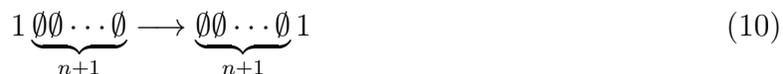
\be\la{9}
\underbrace {\emptyset \emptyset \cdots \emptyset}_{n+1}1\longrightarrow
1\underbrace {\emptyset \emptyset\cdots \emptyset}_{n+1}
\ee
\be\la{10}
1\underbrace {\emptyset \emptyset \cdots \emptyset}_{n+1}\longrightarrow
\underbrace {\emptyset \emptyset\cdots \emptyset}_{n+1}1
\ee.
The basic parameters of the model are $ R = 1 - L$, and $ \m = \l-1 $.
As we increase $ \m $  from $ 0 $ to 1, the effect of pushing to the
right increases and that to the left
decreases. For single particles the hopping rates to the
right and left are still $ R $ and $ L $ respectively.
Thus the driving is controlled by $ R $ or $ L $ and the pushing by
$ \m $  or $ \l $.\\
A few special cases are worth mentioning:\\
When $ \m = \l $, the pushing effect to the right and left are
equal and the rates
are
\be\la{11}   r_n = R {1\over {n+1}} \hskip 2cm   l_n = L {1\over {n+1}}. \ee
In this case, the asymmetry is controlled only by driving.\\
When $ \m = 1 $, then we have maximum pushing to the right and no
pushing to the left.
The rates are
$$ r_n = R, \hskip 2cm  l_0 = L , \hskip 1cm l_{n>0} = 0 $$
Single particles can hop to the left, but can not push other particles to
the left.\\
When $ L = 0 $, particles hop to the right only, and by varying
$ \m $ form 0 to 1
we interpolate between TASEP and the drop-push model [24].
Another way to understand the difference of the two sources of asymmetry
in this problem is to note that
the effect of driving can at least at long times be completely removed
by going to an appropriate
frame of reference while that of pushing can not.
To see this, consider the transformation
\be\la{12}
x_i\to x'_i:= x_i-Vt,
\ee
which is a Galilean boost. Actually, this is not an allowed transformation
of our problem, since the $x_i$'s are integers, whereas $t$ is real. However, if
the probability distribution is sufficiently slowly varying
(e.g. at long  times), then one can define
a probability density function with real variables. The master
equation for this function is (to lowest order)
\be\la{13}
{{\partial P}\over{\partial t}}=-(R-L)\sum_i{{\partial P}\over
{\partial x_i}}+{1\over 2}\sum_i{{\partial^2 P}\over{\partial x_i^2}}.
\ee
In this case, the Galilean boost becomes a symmetry of the space--time
being considered. Going to a reference frame moving with the velocity
$
V=R-L,
$
one obtains
\be\la{14}
{{\partial P}\over{\partial t}}={1\over 2}\sum_i{{\partial^2 P}\over
{\partial{x'_i}^2}}.
\ee
The asymmetry due to driving has been removed.
However, the asymmetry due to pushing remains intact
since the boundary condition
does not change under this transformation.\\
The rest of this paper is devoted to the technical details and elaboration of
the above results. In section 2 we prove that equations (\r{2}) and
(\r{3}) actually
give the above mentioned process. In section 3 we apply the coordinate Bethe
ansatz and show that for this highly nonlocal process the
{\it N}-particle $S$ matrix
is still factorizable. We also find the integral representation of the
{\it N}-particle conditional probability distributions. In section 4
we show that
in the two limiting cases $ \m = 1$ and $ \l = 1, $ the conditional
probabilities can be
expressed in closed form as $ N\times N $ determinants. Section 5 is
devoted to the
mean field solution and discussion of the current density relation. In section
6 we calculate the drift and the diffusion rates for the two particle sector
and compare our results with those of the ordinary partially ASEP [22].
Finally in section 7, we discuss the qualitative picture of the phases for
open systems and also discuss the relation with ordinary ASEP in other
updating schemes.

\section{The Master Equation and the Process}

The master equation is (\r{2}) supplemented by the boundary condition (\r{3}).
 For sectors
of low number of particles (e.g. $N$=2,3), one can repeatedly use (\r{3})
to find the rates. For
general sectors we use the following lemma. \\
{\bf Lemma:}
The boundary condition (\r{3}) implies
\bea\la{15}
&P&(x,x+1,\cdots ,x+n-1,x+n,x+n)=\cr &&(1-r_{n+1})P(x,x+1,\cdots
,x+n-1,x+n,x+n+1)+\cr &&r_{n+1}P(x-1,x,\cdots ,x+n-2,x+n-1,x+n),
\eea
\bea\la{16}
&P&(x,x,x+1,\cdots ,x+n-1,x+n)=\cr &&(1-l_{n+1})P(x-1,x,x+1,\cdots
,x+n-1,x+n,x+n)+\cr &&l_{n+1}P(x,x+1,x+2,\cdots ,x+n-1,x+n,x+n+1),
\eea
where $ r_n $ and $ l_n $ are given in (\r{7}) and (\r{8}).\\
The proof of this lemma is almost the same as that given in [23] and will not
be repeated here. This lemma in fact states how to resolve the singularity
in coordinates of a cluster of {\it n} adjacent particles with that of a single
particle from right and left respectively.\\
Consider now the master equation  for a collection of {\it n}
adjacent particles:
\bea\la{17}
&&\qquad\qquad{\partial\over{\partial t}}P(x,x+1,\cdots ,x+n-1)=\cr
&&R\left[\sum_{i=0}^{n-1}r_iP(x-1,x,\cdots ,x+i-2,x+i-1,x+i+1,\cdots ,
x+n-1)\right.\cr
&-&\left.\left( \sum_{i=0}^{n-1}r_i \right) P(x,x+1,\cdots ,x+n-1)\right]\cr
&+&L\left[\sum_{i=0}^{n-1}l_{n-1-i}P(x,x+1,\cdots ,x+i-1,x+i+1,x+i+2,\cdots
, x+n)\right.\cr
&-&\left.\left( \sum_{i=0}^{n-1}l_i \right) P(x,x+1,\cdots ,x+n-1)\right] .
\eea
It is now clear that the master equation (\r{2}) and the boundary
condition (\r{3})
imply the processes (\r{5})-(\r{8}).\\
\section{The Bethe Ansatz Solution}
\subsection{The case of an infinite lattice}
The Bethe--ansatz solution to the master equation (\ref{2}) is
\be\la{18}
P(x_1,\cdots ,x_N;t)=e^{Et}\Psi (x_1,\cdots ,x_N),
\ee
where
\be\la{19}
\Psi (x_1,\cdots ,x_N)=\sum_\sigma A_\sigma e^{i\sigma( {\bf p})\cdot
{\bf x}}.
\ee
Here {\bf x} and {\bf p} denote $n$-tuples of coordinates and momenta,
respectively, the summation runs over the elements of the permutation group,
and $A_\sigma$'s are coefficients to be determined from the boundary
condition (\ref{3}). Inserting (\ref{18}) in (\ref{2}), we have
\be\la{20}
R\sum_{j=1}^N\Psi (x_1,\cdots ,x_j-1,\cdots ,x_N)+L
\sum_{j=1}^N\Psi (x_1,\cdots ,x_j+1,\cdots ,x_N)=(N+E)\Psi (x_1,\cdots ,x_N),
\ee
or
\be\la{21}
\sum_\sigma A_\sigma e^{i\sigma ({\bf p})\cdot {\bf x}}\left[
R\sum_j e^{-i\sigma (p_j)}+L\sum_j e^{i\sigma (p_j)}\right] =
(N+E)\Psi (x_1,\cdots ,x_N).
\ee
From this, one obtains (as one can remove $\sigma$ from the summations in the
parenthesis)
\be\la{22}
E=\sum_j E(p_j),
\ee
where
\be\la{23}
E(p_j)=R e^{-ip_j}+L e^{ip_j}-1.
\ee
The next step is to determine the coefficients $A_\sigma$, so that the
eigenfunction $\Psi$ satisfies the boundary condition (\ref{3}). It is seen
that nothing from the master equation enters this boundary condition. So the
solution to this boundary condition is just what was found in [23], that is
\be\la{24}
A_{\sigma\sigma_i}=S(\sigma (p_i),\sigma (p_{i+1}))A_\sigma ,
\ee
where $\sigma$ is an arbitrary element of the permutation group, and
$\sigma_i$ is the generator which only interchanges $p_i$ and $p_{i+1}$. The
elements of the two--particle scattering matrix,
\be \la{25}
S_{jk}:=S(p_j,p_k) =- {{1-\l e^{ip_k} - \m e^{-ip_j}}\over
{ 1-\l e^{ip_j} - \m e^{-ip_k}}},
\ee
are thus sufficient to calculate the scattering matrix in the general
$N$-particle sector, and the latter factorizes in terms of the former. As
the scattering matrix is just that obtained in [23], the same reasoning
shows that there are no bound states. So we can write the conditional
probability as
\be\la{26}
P({\bf x};t|{\bf y};0)=\int\prod_{j=1}^N{{\d p_j}\over{2\pi}}e^{E({\bf p})t
-i{\bf p}\cdot{\bf y}}\Psi ({\bf x},{\bf p}),
\ee
where the normalization of $\Psi$ is chosen so that the coefficient of
$e^{i{\bf p}\cdot{\bf x}}$ in $\Psi$ is equal to unity. Integrations run
over $0\leq p_j\leq 2\pi$, and the poles of the scattering matrix are
shifted from the integration region through the prescription
\be \la{27}
S(p_k,p_m)\rightarrow S(p_k+i\epsilon ,p_m),\qquad k<m.
\ee
This ensures that in the physical region $(x_j<x_{j+1}\; ,\; y_j<y_{j+1})$
we have
\be \la{28}
P({\bf x};0|{\bf y};0)=\prod_j\delta_{x_j,y_j}.
\ee
Equation (26) is an integral representation for the conditional probabilities.
More explicit information is obtained after calculating the integrals. For
example in the two particle sector equation (\r{26}) is written as
\newpage
\bea\la{29}
P(x_1,x_2;t|y_1,y_2;0)&=&\int{{\d^2p}\over{4\pi^2}}e^{[E(p_1)+E(p_2)]t
-i(p_1y_1+p_2y_2)}\cr
&&\times\left\{ e^{i(p_1x_1+p_2x_2)}-{{1-\l e^{ip_2}-\m e^{-ip_1}}\over
{1+\epsilon -\l e^{ip_1}-\m e^{-ip_2}}}e^{i(p_1x_2+p_2x_1)}\right\}.
\eea
Using the variables $\xi :=e^{ip_1}$ and $\eta :=e^{-ip_2}$, a contour
integration yields
\bea \la{30}
P(x_1,x_2;t|y_1,y_2;0)&=& e^{-2t}\sum_{m,n}\left\{ {{(Lt)^{m+n+y_2-x_2}
(Rt)^{m+n+x_1-y_1}}\over{n!(n+y_2-x_2)!m!(m+x_1-y_1)!}}\right.\cr
&&-\sum_{p,q}{{p+q}\choose p}\l^p\m^q{{(Lt)^{m+n+y_2-x_1+q}
(Rt)^{m+n+x_2-y_1+p}}\over{n!(n+y_2-x_1+q)!m!(m+x_2-y_1+p)!}}\cr
&&\left.\times\left[ 1-{{\l (n+y_2-x_1+q)}\over{Lt}}-{{\m (m+x_2-y_1+p)}
\over{Rt}}\right]\right\} .
\eea
Here all of the summations run from zero to infinity. A simple calculation
shows that in the limit $L=0$, the result of [23] is obtained.
\subsection{ The case of a periodic lattice}
On an infinite lattice, the set of momenta of the eigenfunctions of the
Hamiltonian are continuous. On a finite lattice, however, this set is
discrete. To obtain this set, consider a lattice of $M$ sites, on which
$N$ particles live. Now, another boundary condition should be added [25]
\be\la{31}
\Psi (x_1, x_2, \cdots , x_N)=\Psi (x_2, x_3, \cdots , x_N, x_1+M).
\ee
This means that one cannot unambiguously define the {\it first}
particle: one can interpret the {\it first} particle as the {\it last} one,
provided its coordinate is enhanced by $M$, the period of the lattice. Applying
the boundary condition (\ref{31}) on the eigenfunction (\ref{19}), we have
\bea \la{32}
\sum_{\sigma}A_\sigma e^{i[\sigma (p_1)x_1+\sigma (p_2)x_2+\cdots
     +\sigma (p_N)x_N]}&=&
\sum_{\sigma}A_\sigma e^{i[\sigma (p_1)x_2+\sigma (p_2)x_3+\cdots
     +\sigma (p_N)(x_1+M)]}\cr
&=&  \sum_{\sigma}A_{\sigma\sigma_0} e^{i[\sigma (p_2)x_2+\sigma (p_3)x_3
     +\cdots +\sigma (p_1)(x_1+M)]},
\eea
where
\be  \la{33}
\sigma_0 (p_1, p_2, \cdots , p_N):=(p_2, p_3, \cdots , p_N, p_1).
\ee
This yields
\be\label{34}
A_\sigma =A_{\sigma\sigma_0}e^{iM\s (p_1)}.
\ee
But,
\be \la{35}
\s_0 =\s_1\cdots\s_{N-1}.
\ee
So, using (\ref{24}),
\bea \la{36}
A_{\s\s_0}&=&A_{\s\s_1\cdots\s_{N-1}}\cr
  &=&A_\s S[\s\s_1\cdots\s_{N-2}(p_{N-1}),\s\s_1\cdots\s_{N-2}(p_N)]\cdots
          S[\s (p_1),\s (p_2)]\cr
  &=&A_\s S[\s (p_1),\s (p_N)]S[\s (p_1),\s (p_{N-1})]\cdots
     S[\s (p_1),\s (p_2)].
\eea
Combining this with (\ref{34}), we arrive at
\be \la{37}
e^{-iM\s (p_1)}=S[\s (p_1),\s (p_N)]\cdots S[\s (p_1),\s (p_2)],
\ee
which can be written as
\be  \la{38}
e^{-iMp_k}=\prod_{j\ne k}S(p_k, p_j).
\ee
These are the Bethe equations of the system, the solution of which provides
the allowed set of discrete momenta. Note that the driving parameter enters
only the energy equation, as in (23), and the pushing parameter enters only
the Bethe equations. Denoting $ e^{-ip_k }$ by $ z_k $,
eqs. (\r{38}) can be rewritten as:
\be \la{39}  z_k ^ M = \prod_{j\ne k } { \l z_j^{-1}+\m z_k - 1 \over
1-\l z_k^{-1}-\m z_j}\ee
This system of equations have a symmetry, namely it is invariant under
$ z \longrightarrow z^{-1} , \l \longrightarrow \m $. This means that if
the set $ z^{\alpha}:= \{z_k^{\alpha} \} $ are the quantized momenta
for the ($\l,\m$ ) system, then the set  $ \omega^{\alpha}:=
\{(z_k^{\alpha})^{-1} \}
$ are the quantized momenta for the ($\m,\l$ ) system.
Thus if we know the spectrum of the former system (see eqs.(22) and (23)):
\be \la{40}  E_{\alpha} = \sum_j( R z_j^{(\alpha)} + L (z_j^{(\alpha)})^{-1}
 - 1 ),\ee
then the spectrum of the latter system is also known
\be E'_{\alpha} = \sum_j( R (z_j^{(\alpha)})^{-1} + L (z_j^{(\alpha)})
 - 1 )\ee
In particular this means that part of the analysis of the spectrum of
ordinary ASEP
($L=\m=0$) which has been done by Gwa and Sphon [25], can be applied
to the
drop-push model [24].

\section{Closed Form of the conditional probabilities in the limiting
cases $ \l = 1 $
and $ \m = 1 $ }
In the special cases $\l =0,1$, one can use a determinant ansatz for the
conditional probabilities [22,23]:
\be\label{42}
P({\bf x};t|{\bf y};0)=e^{-Nt}{\rm det}[G({\bf x};t|{\bf y};0)],
\ee
where $G$ is an $N\times N$ matrix with elements
\be \la{43}
G_{ij}({\bf x};t|{\bf y};0)=g_{i-j}(x_i-y_j;t).
\ee
Inserting (\ref{42}) in (\ref{2}), one obtains
\be\label{44}
{\partial\over{\partial t}}G_{ij}(x;t)=R\; G_{ij}(x-1;t)+L\; G_{ij}(x+1;t).
\ee
The equations obtained by the boundary condition (\ref{3}) (for $\l =0,1$)
are
\be\label{45}
\cases{g_{k-1}(x;t)=g_{k-1}(x-1;t)+\beta g_k(x;t),&$\l =0$\cr
       g_{k+1}(x;t)=g_{k+1}(x+1;t)+\beta g_k(x;t),&$\l =1.$\cr}
\ee
Writing (\ref{44}) in the form
\be\label{46}
\dot g_k(x;t)=R\; g_k(x-1;t)+L\; g_k(x+1;t),
\ee
and introducing the $z$-transform
\be \la{47}
\tilde g_k(z,t):=\sum_x z^x g_k(x;t),
\ee
we obtain from (\r{45}) the following
\be\label{48}
\cases{\tilde g_{k-1}(z,t)={\ds{\beta\over{1-z}}}\tilde g_k(z,t),&$\l =0$\cr
    \tilde g_{k+1}(z,t)={\ds{\beta\over{1-1/z}}}\tilde g_k(z,t),&$\l =1$\cr}
\ee
while (\ref{46}) yields
\be\label{49}
\tilde g_k(z,t)=e^{(Rz+L/z)t}\tilde g_k(z,0)
.\ee
We also note that $\tilde g_k(z,t)$ is the $z$-transform of
the one--particle sector probability; so
\be\label{50}
\tilde g_0(z,0)=\sum_x z^x\delta_{x,0}=1.
\ee
Combining (\ref{48}), (\ref{49}), and (\ref{50}), we arrive at
\be\label{51}
\tilde g_k(z,t)=\cases{e^{(Rz+L/z)t}{\ds{\left({\beta\over{1-z}}
    \right)^{-k}}}, &$\l =0$\cr
    e^{(Rz+L/z)t}{\ds{\left({\beta\over{1-1/z}}\right)^k}},&$\l =1.$\cr}
\ee
The parameter $\beta$ drops from the determinant, so that one can set it
equal to an arbitrary number; we set it equal to unity. From this, one
obtains
\be \la{52}
g_k(x;t)=\sum_{{m,n=0}\atop{n-m+x\geq 0}}^\infty (-1)^{n-m+x}
{k\choose{n-m+x}}{{(Rt)^m(Lt)^n}\over{m!n!}},\qquad\l =0,
\ee
and
\be \la{53}
g_k(x;t)=\sum_{{m,n=0}\atop{n-m-x\geq 0}}^\infty (-1)^{n-m-x}
{{-k}\choose{n-m-x}}{{(Lt)^m(Rt)^n}\over{m!n!}},\qquad\l =1.
\ee
Note that
\be \la{54}
g_k^{(\l =1,R,L)}(x;t)=g_{-k}^{(\l =0,L,R)}(-x;t),
\ee
which is a special case of the symmetry under reflection. By this, we mean
that the system
of equations (2) and (3) is invariant under the following transformations:
$$ x_i\longrightarrow - x_{N+1-i}, \hskip 1cm
R\longrightarrow L, \hskip 1cm
\l\longrightarrow \m$$
\section{Steady state of the system on a Ring}
In this section we consider a ring of $N$ sites on which $M$ particles are
hopping.
The steady state of this system is the one in which all the configurations have
equal weights.Thus all the steady state probabilities $
P( x_1, x_2 , ... x_M ) $ are equal to a
constant. Stationarity of this measure is proved by noting that
 $ P( x_1, x_2 , ... x_M ) = constant $,
satisfies both the master equation (\r{2}) and the boundary condition (\r{3}).
Uniqueness of the measure is ensured by connectivity of the process,i.e.
the fact that every
configuration can be reached from any other by a sequence of transitions [26].
In this
state one can calculate  all the correlation functions by simple
combinatorics. If $ n_k $ is the random variable at site $ k $
 which is 1 if it is
occupied and 0 if it is vacant, then it is well known [15] that
\be \la{55}  <n_j > = { M\over N},\hskip 1cm <n_j n_k > =
{ M (M-1)\over N (N-1)},
\hskip 1cm <n_j n_k n_l > = { M (M-1)( M-2)\over N (N-1)( N-2)},
\ \ \ \ {\rm etc.}\ee
In the thermodynamic limit, when $ M\rightarrow \infty $ and $ N\rightarrow
\infty $ with $ M/N = \rho $, this steady state
approaches the uncorrelated steady state given by the mean field solution.
What we want to do in this section is to find the current density relation
for such a steady state. In general one can find the equation for the rate of
change of the average density at site {\it k}, either by going to the
Hamiltonian
formalism and using the equations $ {d\over dt}<n_k> = < [ n_k , H ] >,$ or
by just looking at the process
and determining the various ways in which this density decreases or increases.
We follow this second approach which is more transparent and intuitive.
The rate of change of density of particles can also be written as a continuity
equation,{\it i.e.} ${d\over dt}<n_k> = J_{k-1} - J_{k}$, where $J_k$ is the
current through site $k$. This current is the algebraic sum of a positive
and a negative current
\be  \la{58}  J_k := R J_k^+ - L J_k ^-, \ee
where $J_k^+$ and $J_k^-$ are due to the hopping of particles to the right
and to the left respectively. Due to the pushing effect, both of these currents
are non-local. The explicit expression of $J_k^+$ and $J_k^-$ are:
\bea  \la{59} J_k^+ &=& r_0 < n_{k}(1-n_{k+1})> + r_1 \bigg(
<n_{k-1}n_{k}(1-n_{k+1})> + <n_{k}n_{k+1}(1-n_{k+2}) > \bigg) \cr
&+& r_2 \bigg( < n_{k-2}n_{k-1}n_k (1-n_{k+1})> + < n_{k-1}
n_{k}n_{k+1} (1-n_{k+2})>\cr &+& < n_{k}n_{k+1}n_{k+2}
(1-n_{k+3})>\bigg) + \cdots , \eea
and
\bea \la{60}  J_k^- &=& l_0 < (1-n_{k})n_{k+1}> + l_1
\bigg( <(1-n_k)n_{k+1}n_{k+2}> + <(1-n_{k-1})n_{k}n_{k+1} > \bigg) \cr
&+& l_2 \bigg( <(1-n_k) n_{k+1}n_{k+2}n_{k+3}> + < (1-n_{k-1})
n_{k}n_{k}n_{k+1} n_{k+2}>\cr &+&  < (1-n_{k-2}n_{k-1}
n_{k}n_{k+1})>\bigg)+ \cdots .\eea
A typical term like $ < n_{k-1}n_{k}( 1-n_{k+1})>$
in the expression of $J_k^+$, is in fact the probability of
the configuration 110 on sites $ k-1, k,$ and $k+1$, respectively.
We know from (\r{5}) and (\r{7}) that
this configuration changes with rate $ r_1$ to $ 011$ on the same sites,
hence this term contributes to the current $J_k^+$. A similar interpretation
is true for other term $ < n_{k}n_{k+1}( 1-n_{k+2})>$, {\it etc}.
In the uncorrelated steady state, the above currents are
calculated to be
\be \la{61}
J^+:=\rho (1-\rho )\sum_{n=0}^\infty{(n+1){\rho^n}
\over{1+{\l\over\m}+\cdots +\left({\l\over\m}\right)^n}},
\ee
and
\be \la{62}
J^-:=\rho (1-\rho )\sum_{n=0}^\infty{(n+1){\rho^n}
\over{1+{\m\over\l}+\cdots +\left({\m\over\l}\right)^n}}.
\ee
from which one obtains:
\be \la{63}
J^+=\cases{\rho (1-\rho )(1+{\ds{2\m\over\l}}\rho )+o({\m\over \l})^2
,&$\ \ \ \ \ \ {\ds{\m\over\l}}<<1$,\cr
\rho,& $\ \ \ \ \ \ {\ds{\m\over\l}}=1$,\cr
{\rho\over {1-\rho}}(1+{\ds{\l\over\m}}\rho(\rho-2))+o({\l\over \m})^2,&
$\ \ \ \ \ \ {\ds{\m\over\l}}>>1$,\cr}
\ee
and
\be \la{64}
J^-=\cases{
{\rho\over {1-\rho}}(1+{\ds{\m\over\l}}\rho(\rho-2))+o({\m\over \l})^2,&
$\ \ \ \ \ \ {\ds{\m\over\l}}<<1$\cr
\rho,&$\ \ \ \ \ \ {\ds{\m\over\l}}=1$,\cr
\rho (1-\rho )(1+{\ds{2\l\over\m}}\rho )+o({\l\over
\m})^2,\cr} \ee
Consider $ J^+$:
It is seen that as far as the pushing effect to the right is small
(${\m\over\l}<<1$), the standard mean
field current of the ASEP gets only corrections of the order
${\m\over \l}$. For medium
pushing, when $ {\m\over \l}\approx 1$, the current is exactly equal
to the density and the
interesting point is that contrary to the case of ASEP, even when the lattice
is filled with particles ($ \rho = 1 $), there is a nonzero current
due to pushing.
At very strong pushing (${\m\over \l}>>1$), the current even diverges
when the lattice is
filled .
It is now instructive to consider only the leading terms of the
total current $J_k$
in different regimes. In the steady state, this current is independent
of the  site number k and hence is denoted by $J$.
We find from (\r{63}) and (\r{64}) the following
\be \la{65}
J=\cases{ \rho \bigg( R(1-\rho )-{L\over 1-\rho}\bigg)+o({\m\over \l})
,&${\ds{\m\over\l}}<<1$,\cr
(R-L)\rho,& ${\ds{\m\over\l}}=1$,\cr
\rho\bigg({R\over 1-\rho}-L(1-\rho)\bigg)+o({\l\over \m}),&
${\ds{\m\over \l}}>>1$.\cr}
\ee
{\bf Remark:} Note that our model incorporates only the totally
ASEP as a special case . Therefore when $ L\ne 0 $, the above results
should not coincide with those of the partially ASEP in the limiting case
$ \mu = 0 $.\\
Consider the case $ {\m\over \l}<<1,$ where we have strong pushing to
the left and weak
pushing to the right. It is seen that for $ L>R$ the current
is always negative, which is expectable on physical grounds.
However, for $L<R$,
when the driving force is to the right, the two effects act in opposite
directions. The current is positive as long as $ \rho < \rho_c:=1-
{\sqrt {L\over R}}$. At $ \rho = \rho _c $ the current
vanishes and for $ \rho> \rho _c $, the pushing effect takes
over and the current becomes
negative. \\

\section{Drift and Diffusion Rates in the Two Particle Sectors }
In this section we want to study the interplay of driving and pushing in
the behaviour of two important dynamical quantities, namely the drift and the
diffusion rates. More generally, we study the long time behaviour of the
quantities $ {d\over dt}<X>$ and ${d\over dt}(<X^2>-<X>^2)$. Our
starting point is the exact calculation of conditional probability of the
the two particles being a distance $ x $  apart, given their initial separation
$ y $. Denoted by $ P_r ( x;t|y;0)$, it is given as:
\bea \la{66}
P_r(x;t|y,;0)&=&\sum_{x_2=-\infty}^\infty P(x_2-x,x_2;t|0,y;0)\cr
&=&\int{{\d^2p}\over{4\pi^2}}e^{Et-i{\bf p}\cdot{\bf y}}\cr
&&\times\sum_{x_2}e^{i(p_1+p_2)x_2}\left[ e^{-ip_1x}+S(p_1,p_2)e^{-ip_2x}
\right]\cr
&=&\int{{\d p}\over{2\pi}}e^{[E(p)+E(-p)]t+ip(y)}(e^{-ipx}+
e^{-ip}e^{ipx}),
\eea
where we have used:
\be \la{67}
S(p,-p)=e^{-ip}.
\ee
From these we arrive at
\be\label{68}
P_r(x;t|y;0)=e^{-2t}\left[\I_{y-x}(2t)+\I_{y+x-1}(2t)\right] ,
\ee
where  I$_n$
is the modified Bessel function of order $n$ with the integral
representation
\be \la{69}
{\rm I}_n(u)=\int{{\d p}\over{2\pi}}e^{inp+u\cos p}.
\ee
It is interesting to note that this probability is independent of the
asymmetry- and drift-parameters. Another way to see this result, is to
derive the equation of evolution for $P_r$. To do so, begin from the master
equation (\ref{2}) for two particles. Using the definition of $P_r(x;t|y;0)$
now abbreviated to $ P_r(x)$, we arrive at
\be \la{70}
\dot P_r(x)=P_r(x-1)+P_r(x+1)-2 P_r(x).
\ee
The boundary condition (\ref{3}) is transformed into
\be \la{71}
P_r(0)=P_r(1).
\ee
It is seen that the driving and the pushing parameters are absent in this
equation. The physical explanation for the absence of $ \m $ or $ \l $
is that, when two particles push each other they do not change their
inter-particle distance. This distance increases by one unit with rate
$R+L$
(particle 2 hoping to the right or particle 1 to the left), and decreases
by one unit with the same rate (particle 2 hoping to the left or particle 1
to the right), and since $ R + L $ has been rescaled to unity, the driving
parameters do not
appear in these equations either.
We should stress that this is not the case in more than two-particle
sectors and the probabilities for relative distance in these sectors do
indeed depend on the above parameters.\\
The next quantities we calculate are the average velocities of particle 1 and
particle 2. Note that particles can not overtake each other and that they keep
their initial order at all times.
We have:
\be \la{72}  <x_i> :=\sum_x x P_i(x) \ee
where $ P_i(x)$ is the probability of finding particle $ i $ at site
$ x $ . The master equation
for these probabilities are obtained from (\r 2) and (\r 3),
using the definitions:
\be  \la{73}
P_1(x):=\sum_{x_2=x+1}^\infty P(x,x_2),
\ee
and
\be  \la{74}
P_2(x):=\sum_{x_1=-\infty}^{x-1}P(x_1,x).
\ee
This calculation finally leads to the following equations
\bea\label{75}
\dot P_1(x)&=&R\{ P_1(x-1)-P(x-1,x)+\m P(x-1,x)\} - ( x\longrightarrow x+1)\cr
 &&+L\{ P_1(x+1)+\l P(x+1,x+2)\} - (x\longrightarrow x-1),
\eea
and
\bea\label{76}
\dot P_2(x)&=&R\{ P_2(x-1)+\m P(x-2,x-1)\}- ( x\longrightarrow x+1)\cr
           &&+L\{ P_2(x+1)-P(x,x+1)+\l P(x,x+1)\}-( x\longrightarrow x-1) .
\eea
We have written these equations in this unsimplified form in
order to convey their simple
physical meaning.
In fact they can also be obtained by intuitive reasoning . Consider for example
eq. (\r{75}). The first two terms
in the curly bracket of the first line are due to particle 1 at site $x-1$
hopping
to an already vacant site at  {\it x } and the third term is due to particle
1 at site $ x-1$ hopping to site $x$ and pushing the already present
particle 2 at this
site to the right. Other terms have similar meaning.
One now obtains from (\r{72}) and (\r{75},\r{76}) the following
\be \la{77}
{{\d <x_1>}\over{\d t}}=R-L-\l P_r(1),
\ee
and
\be \la{78}
{{\d <x_2>}\over{\d t}}=R-L+\m P_r(1).
\ee
One can even derive these equations from the beginning by physical reasoning
without using the master equation. For example we know that the
hopping rate of particle 1
to the right is normally {\it R} unless it is one site behind
particle 2 where its hopping rate will be $ R\m$.
These two terms can be combined to give a positive contribution
$ R + ( R\m - R )P_r(1) $ to the average velocity of particle 1. On the
other hand its hopping rate to the left
is normally $ L $ unless particle 2 is exactly one site to
its right, where its hopping
rate becomes $ L + L\l $. These two terms are then combined to give
a negative contribution  $ -L-L\l P_r(1))$ to
the average velocity. Adding these, one obtains (\r{77}).
The same kind of reasoning
can give (\r{78}).
Using the definitions
\be \la{79}  <r>:=<x_2>-<x_1> \hskip 2cm <X>:={1\over 2}(<x_1>+<x_2>)\ee
we find
\be \la{80}
{{\d <r>}\over{\d t}}=P_r(1),
\ee
and
\bea\label{81}
{{\d <X>}\over{\d t}}&=&R-L+{{\m -\l}\over 2}P_r(1)\cr
                     &=&R-L+{{\m -\l}\over 2}{{\d <r>}\over{\d t}}.
\eea
From these, one obtains
\be \la{82}
<X>=<X>_0+(R-L)t+{{\m -\l}\over 2}(<r>-<r>_0),
\ee
where the subscript 0 refers to initial conditions.
From (\ref{68}), and using the asymptotic behaviour of the modified Bessel
functions, one can obtain the asymptotic behaviour of these expectation
values. We have
\be\label{83}
P_r(1)={1\over{\sqrt{\pi t}}}+O(t^{-3/2}).
\ee
Then, we obtain
\be\label{84}
<r>=C+2\sqrt{{t\over\pi}}+O(t^{-1/2}),
\ee
where $C$ is a constant depending on the initial conditions. So,
\be \la{85}
<X>=<X>_0+(R-L)t+{{\m -\l}\over 2}\left( C+2\sqrt{{t\over\pi}}-<r>_0\right)
+O(t^{-1/2}).
\ee
At long times we have:
\be  \la{86}
{{\d <X>}\over{\d t}}=R-L+{{\m -\l}\over{2\sqrt{\pi t}}} +O(t^{-3/2}).
\ee
It is seen that to leading order the drift rate defined as $ V:=
{d\over dt}<X>$ is only controlled by
driving, and  pushing has only a sub-leading effect.
To calculate the diffusion rate we proceed as follows:
\be  \la{87}
{{\d (<X^2>-<X>^2)}\over{\d t}}={\d\over{\d t}}\left({1\over 2}<x_1^2>+
{1\over 2}<x_2^2>\right) -{{\d (<X>^2)}\over{\d t}},
\ee
where
\be  \la{88}
<x_i^2>=\sum_x x^2\; P_i(x),\qquad\qquad i=1,2.
\ee
Using (\ref{75}) and (\ref{76}), we arrive at
\be  \la{89}
{{\d <x_1^2>}\over{\d t}}=1+2(R-L)<x_1>+\l (L-R)P_r(1)-2\l\sum_x x\;
P(x,x+1),
\ee
and
\be \la{90}
{{\d <x_2^2>}\over{\d t}}=1+2(R-L)<x_2>+\m (R-L)P_r(1)+2\m\sum_x x\;
P(x,x+1).
\ee
From these we obtain
\bea\label{91}
{{\d (<X^2>-<X>^2)}\over{\d t}}&=&1-(\m -\l )<X>P_r(1)+{{1+(\m -\l)(R-L)}\over
2}P_r(1)\cr &+&{{\m -\l}\over 2}\sum_x(2x-1)P(x-1,x)
\eea
The value of the quantity $\sum_x(2x-1)P(x-1,x)$ is calculated
in the appendix . Inserting its value
in the above formula, using the asymptotic form of the modified
 Bessel function,
one obtains
\be\label{93}
\Delta :=\lim_{t\to\infty}{{\d (<X^2>-<X>^2)}\over{\d t}}=1+(\l -\m )^2
\left({1\over 2}-{1\over\pi}\right) .
\ee
We see that pushing has a leading effect on the diffusion rate. The physical
explanation behind this is that although the average distance
between the particles
grows with time as $ t^{1\over 2}$ (see (\r{84})), the width of
the probability distribution (wave packet) of each
particle increases also with rate $ t^{1\over 2}$, thus
the wave packets always overlap and there is always a finite probability that
the particles push each other.

\section{Discussion and Summary}
We have defined a generalized asymmetric exclusion process
with random sequential
updating in which particles besides hoping randomly to the
left and right can also push
their neighboring particles with rates depending on the
 number of these particles.
We have shown that this model, although governed by a
very non-local Hamiltonian of
the spin  chain type, is exactly solvable via the coordinate Bethe ansatz. (For
the type of the Hamiltonian see [23]). Throughout the paper we
have tried to study
the interplay between the two sources of asymmetry, the one due to
driving and the one due to pushing.
Due to the pushing effect, in our model, cluster of particles can
also hop to right
and left. This is similar to what happens in  sequential
updating schemes [20,21], where in one complete updating of the lattice,
clusters of particles move.
However, we remark  that here we have this effect in continuous time
and not discrete time. Specially, in sequential updating scheme, when the
total updating operator is the product of local updating operators,
the probability of hopping
of clusters in one complete update, turns out to be a power of the
hopping rates of single particles
while in our model this is not so.  That is why our Hamiltonian
is very non-local. \\
The fact that despite this non-locality the system
 has an exact solution and its {\it S}-matrix is factorizable
is interesting.
What remains to be done for this model is to study its steady state
( particularly in the
totally asymmetric case {\it L}=0) and its phase structure on open systems when
particles are injected and extracted at the open ends, to see how the simple
phase diagram of the ASEP will be modified due to pushing. Again, due
to the non-locality
of the Hamiltonian of the process, the conventional technique of Matrix Product
Ansatz can not be applied to this problem. Qualitative pictures may be obtained
along the work of ref. [27]. However, one should first decide as to how
to add boundary terms to
this process, that is if particle arrive only at the already
vacant boundary site
or else, they can also push a cluster of particles of arbitrary size,
already present there.
If this is so, then there will be no longer a genuine difference
between the boundary terms and the
bulk.
We believe that, due to these complications and the nonlocal character
of the process, this problem deserves a separate
study.

\section{Appendix}
To calculate the quantity $\sum_x(2x-1)P(x-1,x)$ needed
in section 6, we proceed as
follows:
\bea  \la{93}
\sum_x(2x-1)P(x-1,x)&=&\int{{\d^2p}\over{4\pi^2 i}}e^{Et-i{\bf p}\cdot
{\bf y}}\left[\left({\partial\over{\partial p_1}}+
{\partial\over{\partial p_2}}\right)\sum_xe^{i(p_1+p_2)x-ip_1}\right.\cr
&&\left. +S_{12}\left({\partial\over{\partial p_1}}+
{\partial\over{\partial p_2}}
\right)\sum_xe^{i(p_1+p_2)x-ip_2}\right]\cr
&=&\int{{i\d^2p}\over{2\pi}}\delta (p_1+p_2)\left[ e^{-ip_1}
\left({\partial\over{\partial p_1}}+{\partial\over{\partial p_2}}\right)
e^{Et-i{\bf p}\cdot{\bf y}}\right.\cr
&&\left. +e^{-ip_2}\left({\partial\over{\partial p_1}}+{\partial\over
{\partial p_2}}\right)\left( S_{12}e^{Et-i{\bf p}\cdot{\bf y}}\right)\right]\cr
&=& e^{-2t}\Big\{ (y_1+y_2)[\I_y(2t)+\I_{y-1}(2t)]\cr
&&+t(R-L)[\I_{y+1}(2t)+\I_y(2t)+\I_{y-1}(2t)+\I_{y-2}(2t)]\cr
&&+(\m -\l )\sum_{k=0}^\infty [\I_{y+k}(2t)+\I_{y+k+1}(2t)]\Big\} .
\eea
Using the identity
\be  \la{94}
\sum_{n=-\infty}^\infty I_n(2t)=e^{2t},
\ee
we arrive at
\bea \la{95}
\sum_x(2x-1)P(x-1,x)&=& e^{-2t}\{ (y_1+y_2)[\I_y(2t)+\I_{y-1}(2t)]\cr
&&+t(R-L)[\I_{y+1}(2t)+\I_y(2t)+\I_{y-1}(2t)+\I_{y-2}(2t)]\}\cr
&&+(\m -\l )\Big\{ 1-e^{-2t}\sum_{n=0}^{y-1}[\I_n(2t)+\I_{n+1}(2t)]\Big\} .
\eea

\noindent{\bf Acknowledgement}

\noindent M. Alimohammadi would like to thank the research council of the
University of Tehran and Institute for Studies
in Theoretical Physics and Mathematics, for partial financial support.

\end{document}